\documentclass{article}
\usepackage{spconf,amsmath,graphicx}
\usepackage{algorithmic}
\usepackage{multicol}
\usepackage{graphicx}
\usepackage{textcomp}
\usepackage{xcolor}
\usepackage{float}
\usepackage{stfloats}
\usepackage{booktabs}
\usepackage{makecell, tabularx}
\usepackage{caption}
\usepackage{citesort}
\usepackage{subcaption}
\usepackage[utf8]{inputenc}
\usepackage[ruled,vlined]{algorithm2e}
\title{Learning-Based UE Classification in Millimeter-Wave Cellular Systems With Mobility}
\name{%
	Dino Pjanić$^{\star \dagger}$,\thanks{This work is partially sponsored by the Swedish Foundation for Strategic Research and Ericsson AB.}
	Alexandros Sopasakis$^{\ddagger}$,
	Harsh Tataria$^{\star}$,
	Fredrik Tufvesson$^{\dagger}$, and 
	Andres Reial$^{\star}$
}
\address{%
	$^{\star}$Ericsson AB, Lund, Sweden \\%
	$^{\dagger}$Department of Electrical and Information Technology, Lund University, Lund, Sweden \\%
	$^{\ddagger}$Department of Mathematics, Lund University, Lund, Sweden \\%
	e-mail: \{dino.pjanic, harsh.tataria, andres.reial\}@ericsson.com, \\alexandros.sopasakis@math.lth.se, and fredrik.tufvesson@eit.lth.se}

\begin{document}
	
	\maketitle
	
	\begin{abstract}
		
		Millimeter-wave cellular communication requires beamforming procedures that enable alignment of the transmitter and receiver beams as the user equipment (UE) moves. For efficient beam tracking it is advantageous to classify users according to their traffic and mobility patterns. Research to date has demonstrated efficient ways of machine learning based UE classification. Although different machine learning approaches have shown success, most of them are based on physical layer attributes of the received signal. This, however, imposes additional complexity and requires access to those lower layer signals. In this paper, we show that traditional supervised and even unsupervised machine learning methods can successfully be applied on higher layer channel measurement  reports in order to perform UE classification, thereby reducing the complexity of the classification process.
	\end{abstract}
	
	\vspace{-10pt}
	\section{Introduction}
	\label{Introduction}
	\vspace{-5pt}
	In wireless communication systems, optimization of the radio access network (RAN) has always been an important area \cite{TATARIA1,SHAFIx}. In the context of fifth-generation (5G) systems, a key issue from a network performance viewpoint is how the RAN can adapt to the dynamic radio environment \cite{SHAFI1}. Irrespective of the deployment mode of 5G systems, it is understood that massive multiple-input multiple-output (MIMO) is expected to be the workhorse of its RAN front-haul at the cellular base stations (BSs) \cite{SHAFI1}. In the 24.250 - 52.6 GHz band, analog or digital beamforming in both azimuth and elevation domains is typically used to increase the received signal levels. The beamforming procedure provides access to control and payload signals for network-level decision making. By studying the characteristics of these signals to/from each UE, \emph{spatial fingerprints} unique to each UE can be found \cite{MIMOperformance}. 
	In the related literature, spatial fingerprinting has been used in conjunction with classical machine learning (ML) methods for UE localization via the learned ``features" of the environment \cite{PhasedPos}, or by direct matching \cite{CNN_pos} without learning the environment features. The authors in \cite{ALI1} train a six-layer fully connected network on real-time observations at sub-6 GHz bands to predict beamforming weight coefficients and blockages, while the study of \cite{BURGHAL1} demonstrates the use of a simple, feed-forward neural network for band assignment to different UEs. While \cite{BURGHAL2} surveys an extensive list of related literature, the vast majority of the works in the literature only consider physical layer (PHY) properties of the transmitted/received signal and do not capture the interaction of the PHY with the data link and media access control layers of the system. In reality, these higher system layers greatly manipulate the PHY signals seen to/from the phased array ports which capture the physical amplitude and phase properties, as well as embed protocol level detail for RAN performance optimization.\\ \indent Considering a cross-layer perspective, this paper investigates whether classical ML methods are capable of classifying UEs into different groups by simultaneously processing layer 2 (L2) uplink channel state information-reference signals (CSI-RSs). A key consideration in our analysis is that of UE mobility, thus making the CSI-RSs time varying for each UE. We consider different mobility patterns of the UEs in a typical urban setting and include movements such as walking, cycling, and traveling with cars or public transportation. The objective is to investigate the use of learning algorithms towards UE classification based solely on the reported narrow beams and corresponding Reference Signal Received Power (RSRP) values in dynamic millimeter-wave (mmWave) scenarios. Previous studies on beam management have mostly focused on best beam predictions \cite{MasterThesisBestNarrowBeamOnly} and have not dealt with combining RSRP measurements across multiple wide beams and the corresponding narrow beams \cite{MasterThesisOneWideBeamOnly}.
	In particular, we 
	employ both supervised and unsupervised methods such as tSNE \cite{tSNE} and K-means clustering \cite{BURGHAL2}, combined with principal component analysis (PCA) in order to classify the UE types. Furthermore we train a number of decision tree classifiers in order to further explore whether it is indeed possible to characterize the modality (i.e. pedestrian, car etc) from the measurement reports. 
	
	Our analysis here assumes line-of-sight (LOS) propagation at a center frequency of 28 GHz across a 100 MHz bandwidth. There are several reasons for this: First, one of our objectives is to assess whether unsupervised learning is able to detect different UE types given their individual time evolving CSI-RS reference signals. Second, the majority of the existing studies (see earlier references) confine their focus to PHY parameters and do not consider the effects of PHY-higher layer co-design since it is difficult to decouple the PHY effects from higher layers once certain access control and protocol-level decisions are made. 
	Finally, the results can be useful for approximating the achievable performance obtained by very sparse mmWave channels, since the consensus  is that there only seems to be a few \emph{dominant} multipath components which are active \cite{TATARIA1,SHAFIx}. We note that since our target is to study UE classification at standardized mmWave frequencies, bands outside the one mentioned above is not within the direct scope of the study and hence conclusions made here can not be extrapolated to other frequencies.  
	\vspace{-10pt}
	\section{Simulated System Description and Methodology}
\begin{figure*}
		\begin{center}
			\includegraphics[width=\textwidth,height=4.3cm]{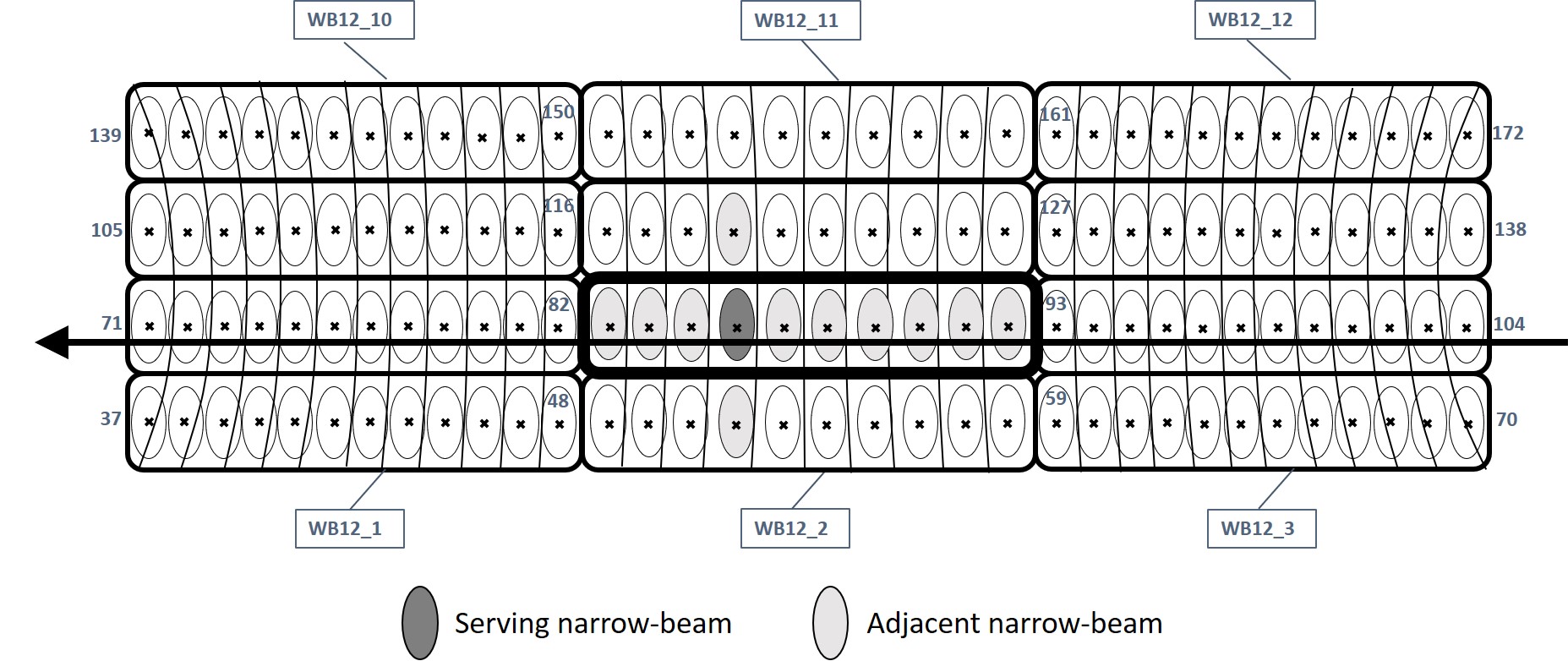}
			\caption{Generated GoB with fixed wide/narrow beam numbering. Conceptual overview of beam refinement procedure: During a typical P2 procedure and with the help of incoming UE measurement reports, NB tracking is performed within the current WB and if RSRP of the best NB is higher than the currently serving NB, a beam switch is initiated. Note: UEs are moving in front of the BS along their lane in same direction as the depicted arrow.}
			\label{fig:Beam_grid}
		\end{center}
	\end{figure*}

	\vspace{-3pt}
	We now describe the commercial grade simulated system which is responsible for generating our data. Unlike previous works, we consider a radio system simulator supporting detailed beam management procedures compliant with standardized 5G systems at mmWave frequencies \cite{38104}. As a result, we are able to simulate a commercial grade 28 GHz phased array antenna module (PAAM) in BS type 1-O configuration containing 192 cross-polarized elements distributed across 8 rows and 12 columns. Analog beamforming capability is modelled with horizontal and vertical inter-element spacings fixed to 0.5 $\lambda$ and 0.7 $\lambda$ with each cross-polarized element having a direction-specific gain pattern given in \cite{36873}. The PAAM is tuned for operation within a bandwidth of 100 MHz (standards compliant). The generated grid-of-beams (GoB) from the PAAM is depicted in Fig.~\ref{fig:Beam_grid} yielding a total of 12 wide beams (WB) and 136 narrow beams (NB). 
	\label{sec:method}
	
\begin{figure}[h!]
		\includegraphics[width=\columnwidth, height=145pt]{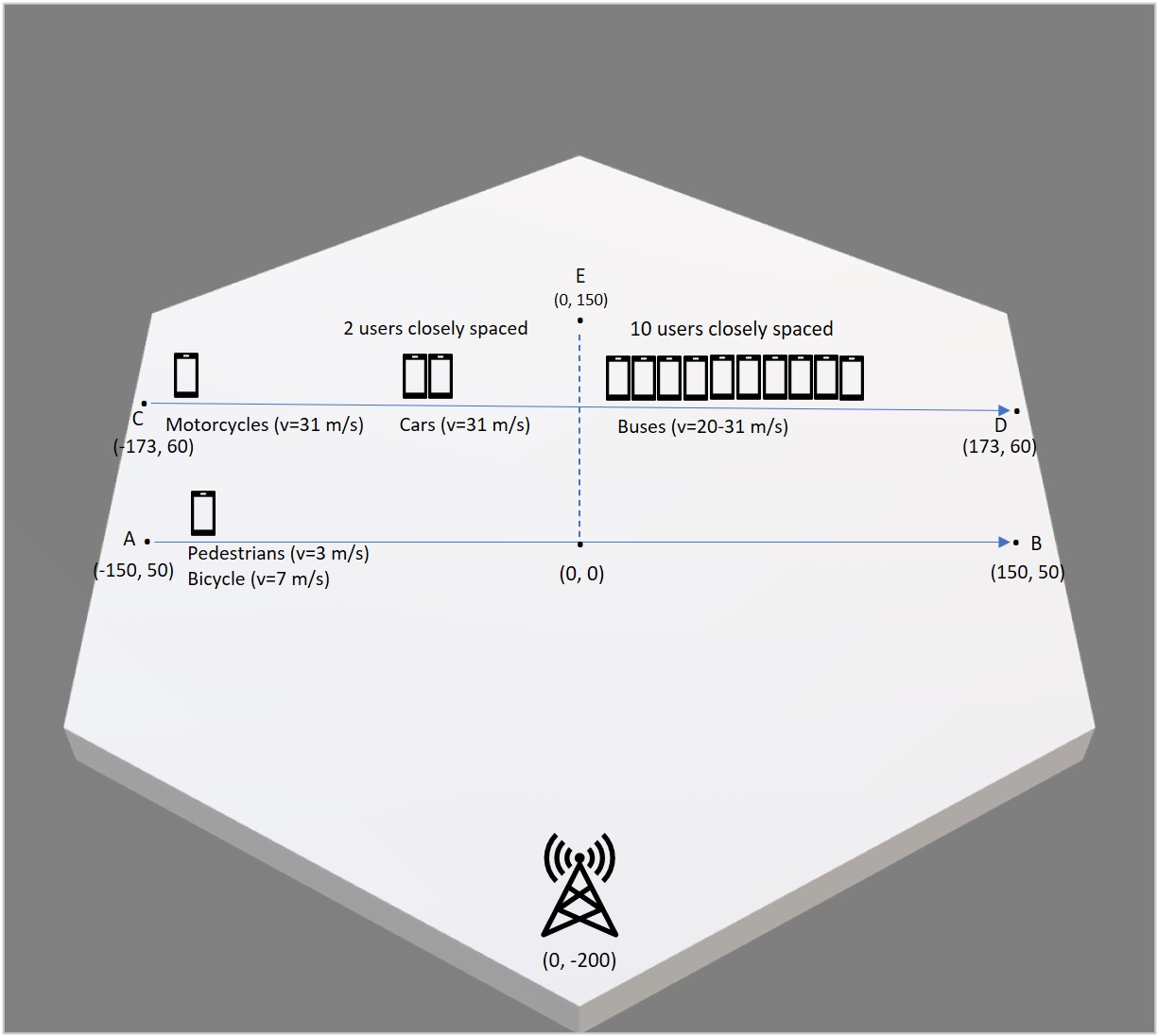} 
		\includegraphics[width=\columnwidth,height=145pt]{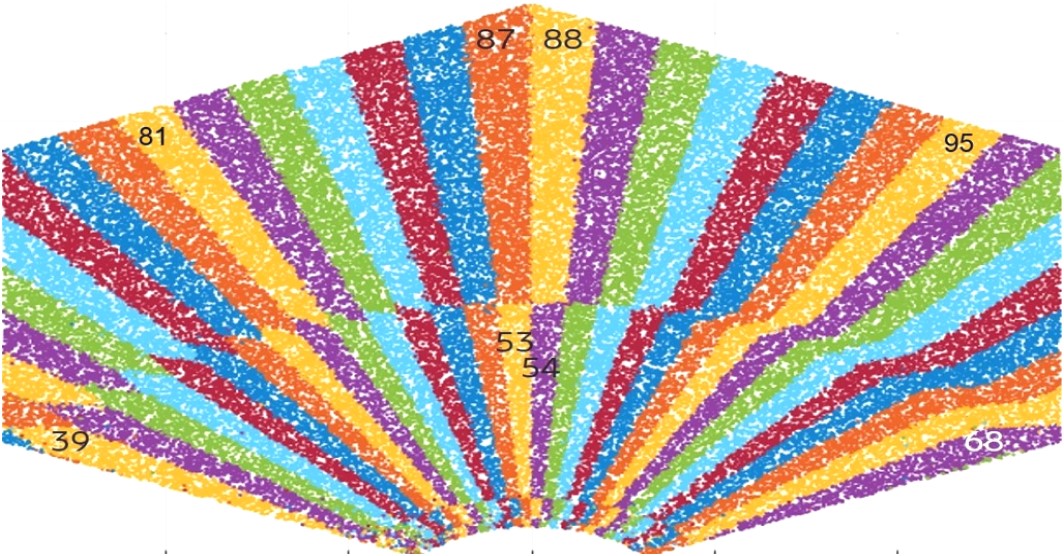} 
		\caption{(Top) Site model. The UEs are placed in LOS moving along two different lanes. (Bottom) An example of narrow-beam coverage showing a subset of beams present in the horizontal plane at a given time instant (as depicted in Fig.~\ref{fig:Beam_grid}).}
		\label{fig:hexagon_site}
	\end{figure}

	\indent The simulated three-dimensional area deploys a single site with one hexagon shaped sector having a radius of 200 m. The BS is deployed in the corner of the hexagon as depicted in Fig.~\ref{fig:hexagon_site}, with height of 30 m having an overall field-of-view of 120$^\circ$ in the azimuth and 40$^\circ$ in elevation. No mechanical downtilt of the BS is assumed. A protection radius of 5 m is kept from the BS periphery where no UEs can be located within 5 m from the BS. 
	Prior to data collection, two UE mobility patterns have been designed, namely slow and fast moving UEs. The first group mimics pedestrian and bicycle while the latter one mimics vehicular UEs such as, motorcycle, car and bus, 5 main classes in total. In addition, user specific mobility patterns have been modelled, presenting a pavement and a street. In order to reduce built-in bias in our data sets, randomness was applied to the starting points as well as UEs movement trajectory. The UEs are moving from left to right in front of the BS having free LOS propagation. Velocities and spacing between UEs are modelled in a way that corresponds to their mobility pattern, as depicted in Fig.~\ref{fig:hexagon_site}. For instance, a car carrying two persons is represented by two UEs physically separated by one meter and moving together at the same time and velocity. The height of each UE is set to 1.5 m. 
	During simulations UEs emerge at predefined starting points along their route and continue moving to the endpoint covering the entire length of the route. In contrast to those, some UEs would emerge at intermediate points and thereby cover just fraction of the entire route. A predefined number of pedestrian UEs is scheduled to cross the street simulating northward movement away from the BS. During the simulation multiple parameter settings are triggered and evaluated for different loads of UEs, and several seeds are used to ensure statistical confidence.
	Fig.~\ref{fig:hexagon_site} shows narrow beam coverage with corresponding numbering/indexing across the simulated geographical area (as generated from the system explained in Fig.~\ref{fig:Beam_grid}). We note the above was done in conjunction with uplink file transfer protocol traffic patterns. 

	\vspace{-10pt}
	\section{Problem Definition and Formulation\label{sec:systemModel}}
	\vspace{-5pt}
	Maintaining good radio link reliability is a key challenge for mmWave communication systems, especially when mobility is incorporated. Directional links, however, require fine alignment of the transmitter and receiver beams, achieved through a mechanism known as beam management \cite{BF_tutorial}. In line with \cite{38802}, three downlink layer 1/2 (L1/L2) beam management procedures, commonly known as P1, P2 and P3 are involved. In this study, our primary focus was on the P2 procedure which handles beam tracking at the BS (a.k.a. gNB). Note that tracking refers to gNB refining beams (e.g., sweeping through all the narrow beams over a small range) where UEs detect the best (service) beam and report its index to gNB. As illustrated in Fig.~\ref{fig:Beam_grid}, the P2 CSI-RS measurements were performed by the moving UEs (as specified in \cite{36133}) for all NB within the same WB (synchronization signal block (SSB) beam) and the reported RSRP measurements were collected.\\
	\indent A considerable amount of literature has been proposing different ML approaches for classification of UEs, mostly applied on data sets generated on physical radio channel attributes such as: e.g., precoding schemes, modulation scheme \cite{Dissertation} or channel covariance \cite{ChannelCovarianceUserClassy}. As UEs tend to move along predefined routes in the physical environment while establishing fingerprints at the BS, e.g. by channel measurement, we argue that using L2 report data-sets can be advantageously utilized in order to simplify the UE classification due its less complex nature compared to physical channel parameters. This approach opens up for new ways of learning from UE mobility patterns and thereby possibly prevent UEs from ending up in unfavorable radio channel conditions.
	\vspace{-12pt}
	\section{Un/supervised Learning Framework\label{sec:problem}}
	\vspace{-7pt}
	To perform our analysis we utilize the fact that mmWave communication is sensitive to UE motion. We explore whether machine learning methods can learn motion characteristics and perhaps even users intent from RSRP measurements as reported on different narrow-beams.
	One difficulty is that our measurement report, as in real life, is typically ill-balanced. Therefore, in the case of K-means, we first perform Principal Component Analysis (PCA) in order to uncover the clustering structure of our data in an unsupervised setting. We use 50 components for the PCA which accounts for 97\% of the cumulative explained variation. 
	In order to ascertain the number of clusters $K$ we employ the elbow method \cite{elbow} and subsequently explore classification of UEs into the main five classes comprising our data: pedestrian, bicycle, car, bus and motorcycle. 
	
	In the subsequent analysis, we explicitly avoid comparing multiple unsupervised classification methods and questions about optimality of one approach over another. Our primary objective is to understand whether it is feasible to utilize knowledge of the time evolving CSI-RS signals for classifying different UEs. Such information can then be used for predicting and optimizing their radio resource requirements. As such, from a ML viewpoint, we opt for the simplest combination of the K-means clustering with PCA, rather than a more complex supervised or unsupervised learning approach based on deep learning methods. We point out that while these are extremely interesting directions, we cover them in detail in a follow up extension of this paper.
	
	\subsection{Feature extraction and the data}
	Each UE measurement report contains 12 RSRP CSI-RS measurements together with their corresponding narrow-beams and arrives on a approximate 40 ms basis. The total number of such reports, which essentially indicates the size of our dataset, is highly dependent on the duration of simulation, as well as the UE velocities as described in Section \ref{sec:method}. Identity information such as the UE identity number, UE location and time are not exposed to the ML algorithm but rather used as labeled references when interpreting results of classification.
	When dealing with incomplete and unbalanced data sets a number of approaches are possible. To counter the data imbalance we employ PCA as a pre-processing step. As explained previously, we apply PCA with 50 components which accounts for 97\% of explained variation in the data. Furthermore, due to data inhomogeneity we apply feature scaling where the range of each feature is normalized so that it contributes proportionately. This approach is essential since the clustering methods calculate distances between data points. Subsequently the two features, RSRP and corresponding narrow-beams, are stacked on top of each other representing one single UE fingerprint at the base station at a certain report time. In tests, not presented here, we observed that having only RSRP or only the narrow beam indices as input to our ML model results in degradation of the classification rate. 
	
	\label{sec:ML_Framework}
	\section{Results and Discussion}
\begin{figure}[h!]
		\begin{center}
			\centerline{\includegraphics[scale=0.63]{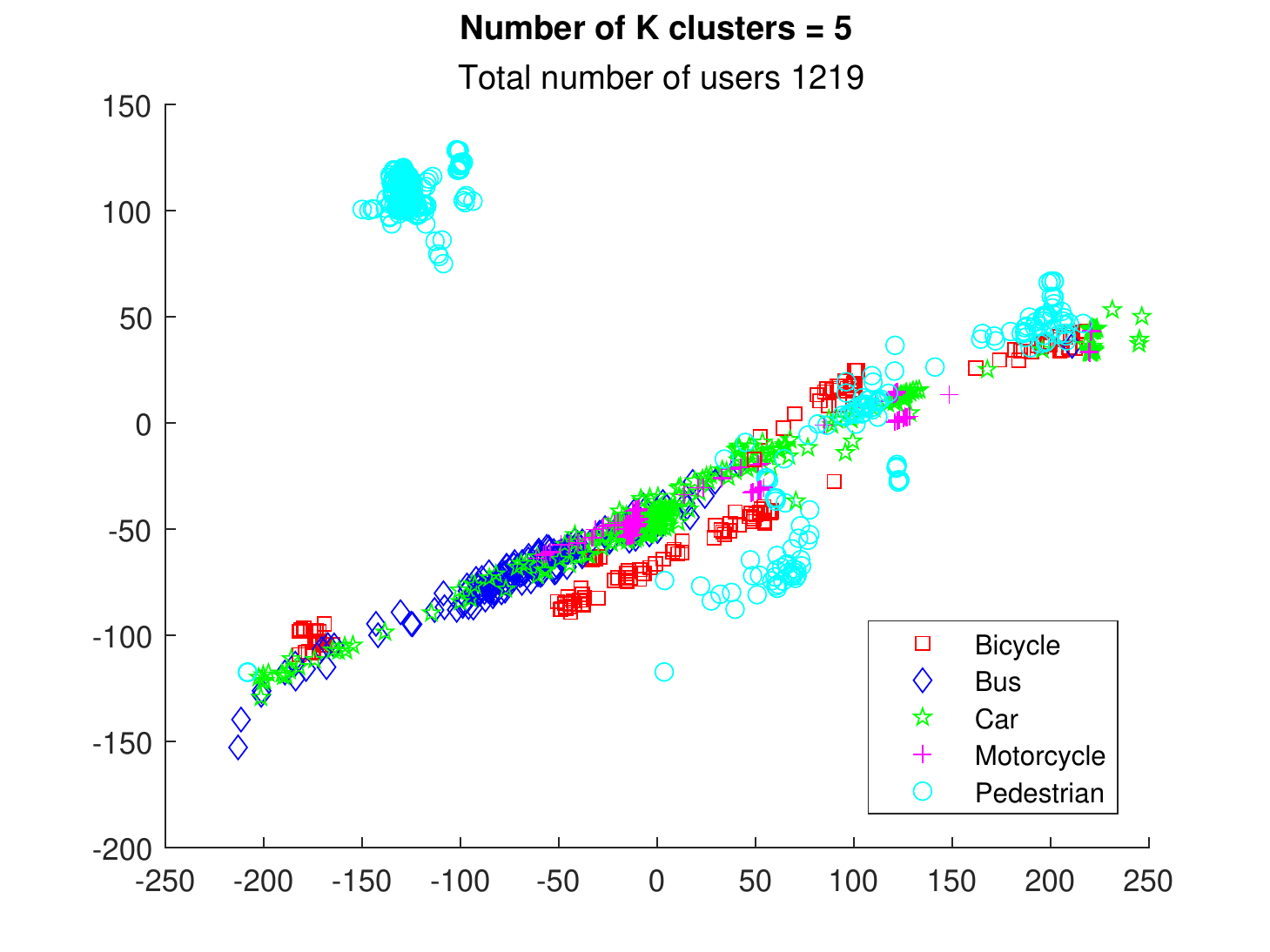}}
			\caption{K-means clustering results for all 5 classes of UEs. Training on 5 groups: 165 bicycles, 180 buses, 320 cars, 130 motorcycles and 424 pedestrians. Class separation is not obvious at this scale but shows pattern similarities may exist between classes of slow movers or classes of fast movers.}
			\label{fig:AllUsersClustered}
		\end{center}
		\vspace{-15pt}
	\end{figure}

\begin{figure}[h!]
		\begin{center}
			\centerline{\includegraphics[scale=0.63]{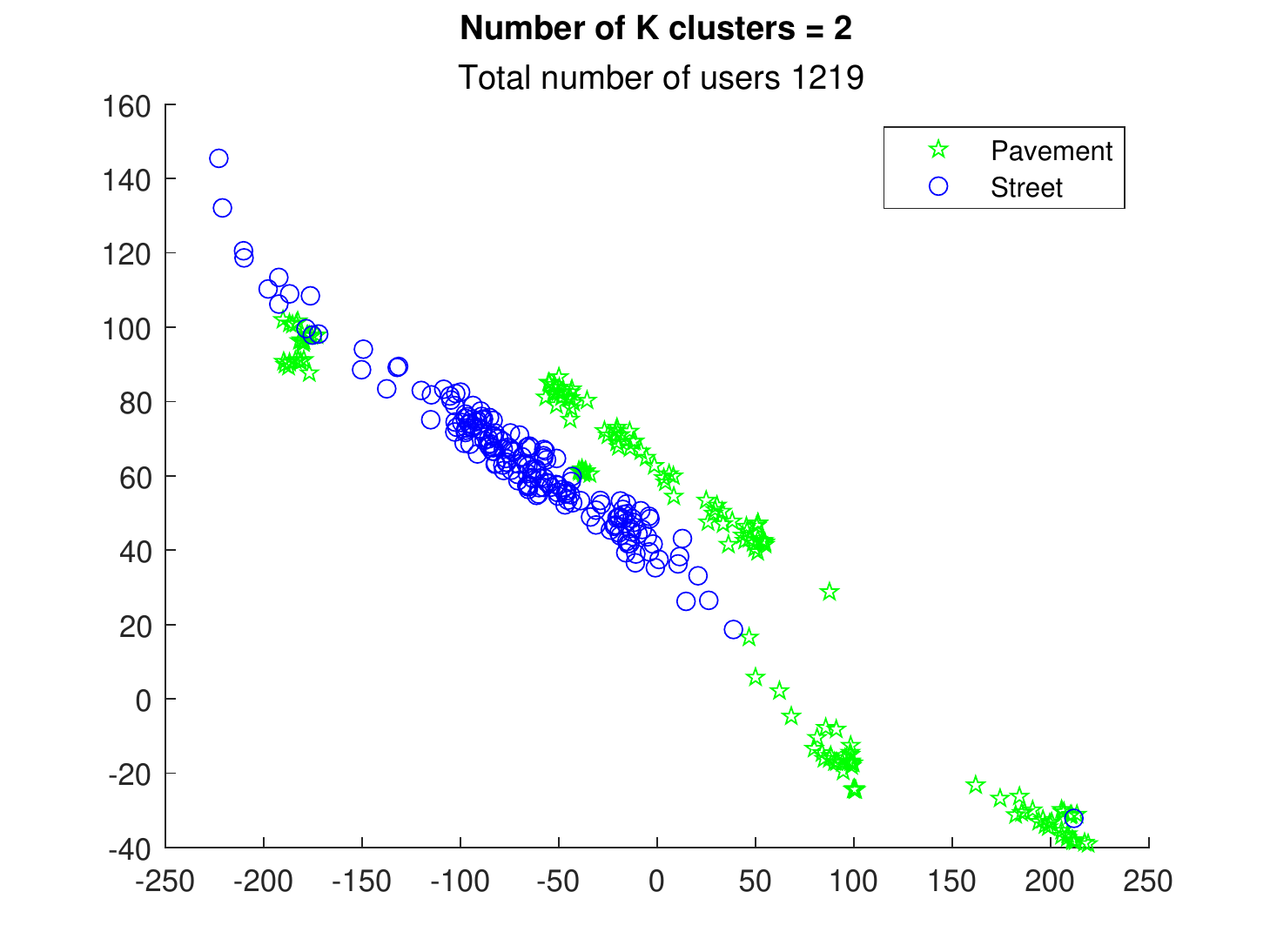}}
			\caption{K-means clustering results for all 5 classes of UEs. Training on 2 groups: fast 630 (cars, busses, motorcycles) versus 589 slow-moving (bicycles, pedestrians) UEs.}
			\label{fig:StreetVsPavement}
		\end{center}
		\vspace{-25pt}
	\end{figure}

	We  now present results from both unsupervised and supervised ML models towards our goal: UE classification from their signals. As discussed in Section \ref{sec:method}, our data consists of 5 main classes, with sub-classes (groups), of UEs according to their mobility pattern. We first train our algorithm to detect all 5 classes. The results are presented 
	in Fig. \ref{fig:AllUsersClustered}. There is significant overlap between the different classes which may not be so surprising given how similar UEs must look for groups such as cars and busses or pedestrians and cyclists. This underscores the need to explore whether we can learn to identify larger groups such as for instance slow and fast movers. We will undertake this task next.
	Using the sama data from Fig. \ref{fig:AllUsersClustered} a similar although perhaps 
	more intriguing result emerges when we set $K=2$ groups. Specifically we observe in Fig. \ref{fig:StreetVsPavement} that some separation exist between fast (cars, busses, motorcycles)  and slow (bicycles and pedestrians) moving UEs. The overlap has now reduced significantly when only grouping into 2 groups. We explore this initial result further below. 
	\subsection{Slow-moving UEs - pedestrians vs bicycles}
\begin{figure}[h!]
		\begin{center}
			\centerline{\includegraphics[scale=0.58]{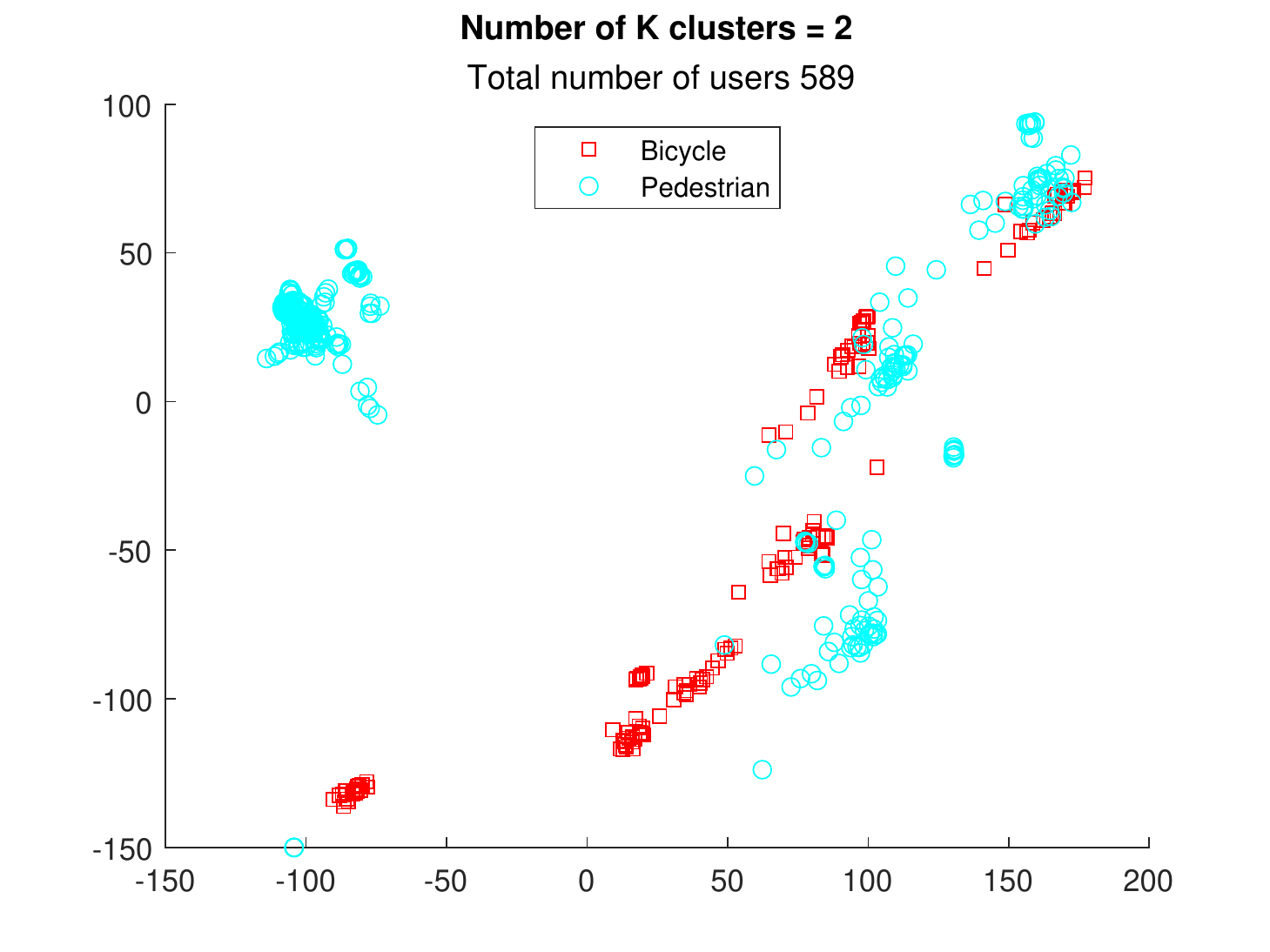}}
			\caption{K-means clustering results for slow-moving UEs. Training on 2 groups: 165 bicycles versus 424 pedestrians. Compare with Fig. \ref{fig:RandomPedestrianMoverVsCrossStreetClassification} where pedestrians are further distinguished into crossing/non-crossing the street. }
			\label{fig:PavementUsersClassification}
		\end{center}
		\vspace{-35pt}
	\end{figure}

\begin{figure}[h]
		\begin{center}
			\centerline{\includegraphics
				[scale=0.58]
				{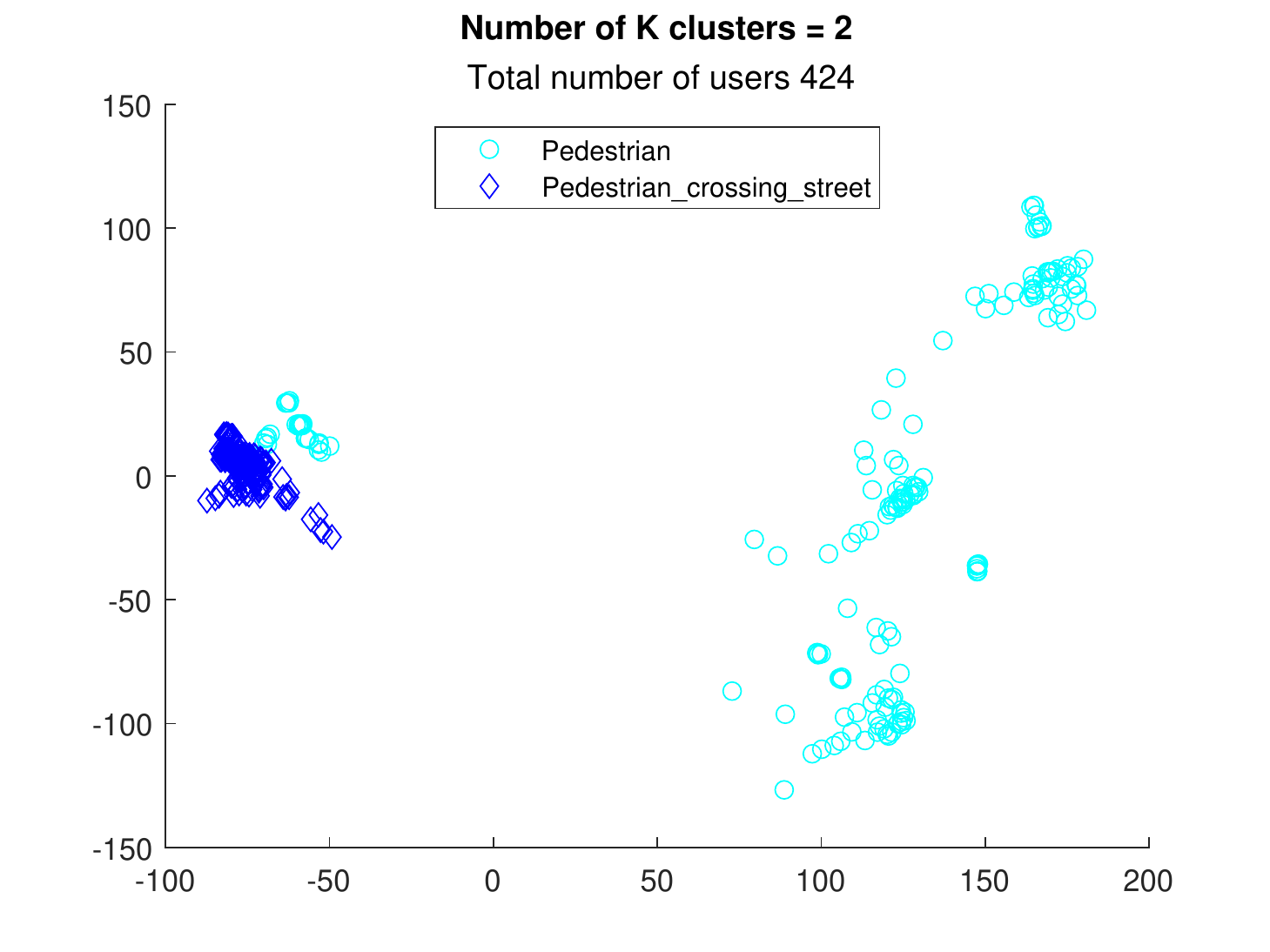}}
			\caption{K-means clustering results after training on just the class of pedestrian UEs in Fig. \ref{fig:PavementUsersClassification}. Training on 2 groups: 182 pedestrians non-crossing the street versus 242 crossing the street. Pedestrians crossing the street seem to be mainly responsible for that single cluster.}
			\label{fig:RandomPedestrianMoverVsCrossStreetClassification}
		\end{center}
		\vspace{-15pt}
	\end{figure}

	It is not surprising, especially based on the baseline results of Fig.\ref{fig:StreetVsPavement}, that velocity of a given UE can be a good indicator as to whether it should be classified as a car or a pedestrian. Differentiating however between similarly moving UEs should be more challenging. 
	We undertake this task in an unsupervised setting for now by examining a mixed group of 589 UEs consisting of 424 pedestrians and 165 bicycles.  We note that UEs in that dataset move along identical trajectories on the pavement with either small or no variation and ending/starting locations unrelated to their group type.  We also note that some of the pedestrians will also cross the street while the bicycles do not. We train on this data with a PCA method followed by K-means as discussed earlier. 
	The results in Fig.~\ref{fig:PavementUsersClassification} show a rather clear separation between the two classes of users with a surprisingly small overlap even though the 2 groups have almost similar moving velocities.
	Specifically, we observe a clear diagonal linear yet separated trend in the majority of the data - we should point out, that the clustering space is non-dimensional and has no physical meaning. We also note an interesting cluster of pedestrians at the top left corner of that figure which seems to defy the overall pattern. To better understand this cluster we then train our algorithm on the single group of 424 pedestrians from Fig.~\ref{fig:PavementUsersClassification} while requiring that $K=2$ once again. The results presented in Fig.~\ref{fig:RandomPedestrianMoverVsCrossStreetClassification} verify that the majority of that sub-group is attributed to the street-crossing pedestrians. The results therefore show that identification and separation among all classes and even sub-group of the slow UEs could be possible.
	\vspace{-25pt}
	\subsection{Classification\label{classificationsection}}
\begin{figure}
		\begin{center}
			\centerline{\includegraphics[scale=0.43]{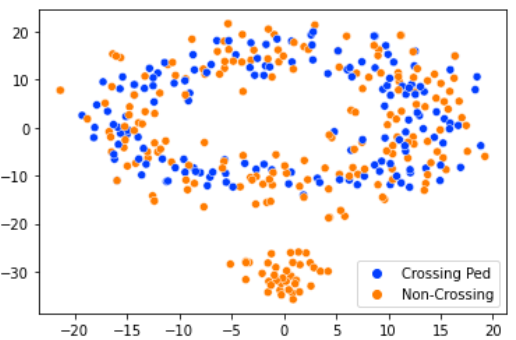}}
			\caption{tSNE clustering results. 242 pedestrians crossing street vs 182 non-crossing. As expected, due to short (40ms) reporting the non-crossing are miss-identified.}
			\label{fig:tSNE}
			\vspace{-30pt}
		\end{center}
	\end{figure}

	We now train a number of different decision tree type (supervised) classifiers in order to identify specific UE mobility classes based on this data. Furthermore we task the classifiers to distinguish modality with high probability (see Table \ref{table:classfastvsslow}) based on only a short (40 ms) sequence of RSRP and narrow-beam numbers reported from each UE.
	We present a list of the most successful of those and the respective type of UE data used for their training in Table \ref{table:classfastvsslow}. Our main metric for success or failure here is the miss-classification rate on unseen data. The best classifier, the Extra Trees Regressor, achieves a miss-classification rate of $2\%$.
	\begin{table}[h]
		\begin{center}
			\begin{tabular}{ccc}
				\hline
				Classifier & Data used & Miss-Classif \%.\\
				\hline
				Extra Trees Regres.& Ped-nc, Car, MC & 1.8\\
				Extra Trees Regres.& Ped-cr, Car, MC & 2.2\\
				Extra Trees Regres.& Ped, Car, MC & 5.2\\
				Ada Boost Regres.& Ped, Car, MC & 8.1
			\end{tabular}
		\end{center}
		\caption{Decision trees and respective miss-classification rates. Data distribution used: All pedestrians 424 (Ped), 242 Pedestrians crossing (Ped-cr) and 182 non-crossing (Ped-nc). 320 Cars and 130 MC$=$Motorcycles.
		}
		\label{table:classfastvsslow}
	\end{table}
	Overall we found that training a classifier to distinguish the pedestrians crossing the road from those not crossing (see also clustering Fig. \ref{fig:tSNE}) was the most difficult task. That however was expected since we essentially requested an impossible task from the classifier. The fraction of pedestrians who eventually cross the street do so only for a brief part of their motion and until that moment they would be indistinguishable from pedestrians intending not to cross the street. Thus the classifier, based on the information it receives, correctly identifies those pedestrians as part of the non-crossing group. It is therefore not surprising that the classifier was not able to detect that particular sub-group of UEs based on the short (40 ms) history of the training sequence provided. 
	\section{Conclusions and Future Work}
	We identified ways of classifying UEs in dynamic millimeter wave scenarios by employing conventional ML techniques on CSI-RS measurements. 
	This initial study in clustering and classification of UEs based on their network measurement reports alone, without considering physical positioning or other supporting information, shows that it is possible to infer the mobility mode of the UEs with some success. The work presented here offers insights towards new beam prediction mechanisms in mobility-aware MIMO scenarios. These results together with trajectory forecasting (future research) could in turn provide useful information when preparing hand-over and expected resource demands in order to ease and avoid operational bottlenecks. 

	\bibliographystyle{IEEEbib}
	
\end{document}